\documentclass[9pt,Preprint, dvipsnames]{lapreprint}

\usepackage[version=4]{mhchem} 
\usepackage{siunitx}    
\usepackage{pdflscape}  
\usepackage{rotating}   
\usepackage{textgreek}  
\usepackage{gensymb}    
\usepackage[misc]{ifsym} 
\usepackage{orcidlink}  
\usepackage{listings}   
\usepackage{colortbl}   
\usepackage{tabularx}   
\usepackage{longtable}  
\usepackage{subcaption}
\usepackage{multirow}
\usepackage{snotez}     

\usepackage{annotate-equations}

\usepackage{xcolor}
\usepackage{algorithmic}
\usepackage{graphicx}
\usepackage{textcomp}
\usepackage{listings}
\usepackage{booktabs}
\usepackage{bold-extra}
\usepackage{tcolorbox}
\usepackage{enumitem}
\usepackage{hyperref}
\usepackage{adjustbox}
\usepackage{comment}
\usepackage[utf8]{inputenc}
\usepackage{xspace}
\usepackage{balance}
\usepackage[frozencache,cachedir=.]{minted} 
\usepackage{xcolor}

\DeclareSIUnit\Molar{M}

\usepackage[			
	backend=biber,      
    style=numeric,   
	natbib=true,		
	hyperref=true,	    
	alldates=year,      
    uniquename=false,   
    maxbibnames=99,     
]{biblatex}

\AtEveryBibitem{
    \clearfield{urlyear}
    \clearfield{urlmonth}
    \clearlist{language}
}

\newtcolorbox{rqbox}[1][]{
    colback=gray!10,
    colframe=gray,
    arc=1mm,
    boxrule=0.5pt,
    coltitle=black,
    fonttitle=\bfseries,
    title=#1
}

\def\BibTeX{{\rm B\kern-.05em{\sc i\kern-.025em b}\kern-.08em
    T\kern-.1667em\lower.7ex\hbox{E}\kern-.125emX}}

\newcommand{\comet}{\textsc{Comet}}
\newcommand{\bleu}{\textsc{bleu}}
\newcommand{\rogue}{\textsc{rogue}}
\newcommand{\rougel}{\textsc{rogue-l}}
\newcommand{\meteor}{\textsc{meteor}}
\newcommand{\abst}{\textsc{ast}}
\newcommand{\cfg}{\textsc{cfg}}

\newcommand{\pdg}{\textsc{pdg}}
\newcommand{\cpg}{\textsc{cpg}}
\newcommand{\cnn}{\textsc{cnn}}
\newcommand{\dl}{\textsc{dl}}
\newcommand{\ie}{\textit{i.e.,}}
\newcommand{\eg}{\textit{e.g.,}}
\newcommand{\etal}{\textit{et al.}}

\newcommand{\github}{\textsc{GitHub}}
\newcommand{\api}{\textsc{api}}

\newcommand{\chatgpt}{\textsc{ChatGPT}}
\newcommand{\deltagraph}{\textit{delta graph}}

\newcommand{\ml}{\textsc{ml}}

\newcommand{\nbc}[3]{
		{\colorbox{#3}{\bfseries\sffamily\scriptsize\textcolor{white}{#1}}}%
		{\textcolor{#3}{\sf\small$\blacktriangleright$\textit{#2}$\blacktriangleleft$}}}

\title{COMET: Generating Commit Messages using Delta Graph Context Representation
}

\author{Abhinav Reddy Mandli}
\author{Saurabhsingh Rajput}
\author{Tushar Sharma}

\affil{Dalhousie University, Canada}

\correspondence{saurabh@dal.ca}{}

\data[]{Dataset is made available on \href{https://doi.org/10.5281/zenodo.7902315}{Zenodo}.}
\funding[]{The authors are partially supported 
by the Natural Sciences and Engineering Research Council of Canada (NSERC) 
through grant NSERC Discovery RGPIN/04903.}

\leadauthor{Mandli}
\shorttitle{Generating Commit Messages using Delta Graph Context Representation}

\begin{document}

\maketitle

\begin{abstract}
Commit messages explain code changes in a commit and facilitate collaboration among developers.
Several commit message generation approaches have been proposed; however, they exhibit limited success in capturing the context of code changes. We propose \textbf{\comet{} (\underline{C}ontext-Aware C\underline{o}mmit \underline{Me}ssage Genera\underline{t}ion)}, a novel approach that captures context of code changes using a graph-based representation and leverages a transformer-based model to generate high-quality commit messages.
Our proposed method utilizes \deltagraph{} that we developed to effectively represent code differences. 
We also introduce a customizable quality assurance module to identify optimal messages, mitigating subjectivity in commit messages. 
Experiments show that \comet{} outperforms state-of-the-art techniques in terms of \bleu-norm{} and \meteor{} metrics while being comparable in terms of \rougel{}.
Additionally, we compare the proposed approach with the popular gpt-3.5-turbo model, along with gpt-4-turbo\textemdash{}the most capable GPT model,
over zero-shot, one-shot, and multi-shot settings.
We found 
\comet{} outperforming the GPT models, 
on five and four metrics respectively and provide competitive results with the two other metrics.
The study has implications for researchers, tool developers, and software developers.
Software developers may utilize \comet{} to generate context-aware commit messages.
Researchers and tool developers can apply the proposed \deltagraph{} technique in similar contexts, like code review summarization. 
\end{abstract}

\section{Introduction}

Commit messages are essential in documenting 
changes carried out in each commit
to maintain and understand code evolution~\parencite{Hindle2009, barnett2016, rebai2020}.
Writing good commit messages promotes effective communication and seamless collaboration within a software development team~\parencite{tao2012,huang2014}. 
Meaningful commit messages ease the burden of many tasks, including debugging and software lifecycle management~\parencite{yan2016, hassan2008}.
A good quality commit message is succinct yet clearly articulated and informative to convey the nature and purpose of the change~\parencite{agrawal2015, cortes2014}, \ie{} the \textit{why} aspect,
as well as the summary of the changes made to the code 
\ie{} covering the \textit{what} aspect~\parencite{tian2022}, along with the context of the code change.

\textit{Context} plays an important role in commit message generation because the effectiveness of the generated message depends on the surrounding context in addition to code changes~\parencite{sillito2008}.
The context refers to the surrounding code in which the change occurs; it provides vital information to understand the rationale behind a change. As shown in Listing~\ref{lst:commit_context}, looking at the diff of changed lines suggests migrating \textsc{http} to \textsc{https} connection. However, the surrounding unchanged code reveals the URL is used to retrieve user data on profile pages. This context helps generate an informative message that the purpose is to \textit{retrieve user data securely}.


\begin{listing}[ht]
\begin{minted}[breaklines]{java}  
// User profile page
public void viewProfile(User user) {

// Retrieve user data
\end{minted}

\begin{minted}[breaklines,highlightcolor=red!30,highlightlines={2-3}]{java} 
- String url = "http://example.com/data";
- HttpURLConnection connection = (HttpURLConnection) url.openConnection();  
\end{minted}

\begin{minted}[breaklines,highlightcolor=green!30,highlightlines={1-4}]{java}
+ SSLContext sslContext = SSLContext.getInstance("TLSv1.2");
+ sslContext.init(null, null, null);
+ HttpsURLConnection connection = (HttpsURLConnection) url.openConnection();
+ connection.setSSLSocketFactory(sslContext.getSocketFactory());
\end{minted}

\begin{minted}[breaklines]{java}
// Render profile view
renderPage(user); 
}
\end{minted}

\begin{minted}[breaklines,highlightlines={1-3}]{java} 
 // Commit message without context: Migrate HTTP connection to HTTPS
 // Commit message with context: Use HTTPS to securely retrieve user data on profile pages
\end{minted}
\caption{Commit message generated with and without code context}
\label{lst:commit_context}
\end{listing}

However, developers often do not provide relevant, good-quality commit messages that align with the commit context. 
According to a study on five popular open-source projects on \github{}~\cite{tian2022}, 
approximately $44\%$ of commit messages lack both \textit{what} and \textit{why} aspects. 
Automated commit message generation approaches attempt to reduce the burden on developers by generating effective commit messages taking into account the performed changes within a context.

Over the years, several approaches have been proposed to generate commit messages automatically.
The initial approaches, such as \textit{ChangeScribe}~\parencite{linares2015} used a code summarization approach to generate commit messages. 
However, such rule-based 
approaches~\parencite{Buse2010,Moreno2017} 
are limited in their ability to capture the context and subjectivity of code changes.
Subsequently, researchers proposed a Neural Machine Translation {\sc (nmt)}-based method to generate commit messages from \textit{code diffs}~\parencite{jiang2017}.
However, these approaches do not fully leverage the syntactic and structural properties of code, and fail to capture the complete context of source code changes. 
Similarly, techniques based on information retrieval~\parencite{Liu2018,huang2020,Wang2021} rely on a reference dataset during inference and are prone to the limitations of information retrieval models, such as the out-of-vocabulary problem~\parencite{Xu2019,dong2022,liu2020}.


Recently, transformer-based large language models (\textsc{LLM}s)~\parencite{feng2020,wang2021codet5} 
have demonstrated impressive capabilities in various code-related tasks, 
such as generating natural language text from code, 
including code comments and commit messages~\parencite{jung2021,shi2022, Sharma2024}.
These models have shown to be effective at capturing complex relationships and patterns in source code by utilizing attention mechanism.
However, the majority of the past approaches that utilize these models do not consider the structural and semantic information of code changes, as they treat code as a linear sequence of tokens.
In addition, their capability to capture the code context in which the changes have been made is yet to be proven. 

In a \ml{}-based method,
the dataset quality significantly influences the training process and the generated outcome.
Previous studies have used various filtering techniques for their datasets based on factors
such as message length, 
code diff length, 
verb direct object dependency, 
and number of classes~\parencite{liu2020, jiang2017}.
However, the \textit{quality of the commit message} itself has not been considered as a metric for filtering in most of the past approaches. 
The poor quality of commit messages that are used in these studies as training samples can significantly impact the performance of these models, leading to substandard results. 
This can be attributed to the "Garbage in, Garbage out" principle~\parencite{sanders2017}, 
where subpar input data leads to inferior output. 
Therefore, it is imperative to use high-quality commit messages in the training phase to improve the performance of the used machine learning (\ml{}) models.

The format of commit messages may vary depending on the context, purpose, and organizational needs.
In practice, varying conventions and templates~\parencite{conventional-commits,qoomon2023} exist to guide developers in structuring their commit messages, 
such as specifying change types including "fix," "feat," or "chore," 
at the beginning of a commit message.
These conventions aid developers in ensuring a standard is followed across their organization. 
It implies that automated commit message generation approaches need to be flexible to accommodate organization-specific message types while still being generalizable enough to cover base scenarios.
The current approaches do not provide this flexibility.

We propose \textbf{\comet{}} (\underline{C}ontext-aware C\underline{o}mmit \underline{Me}ssage Gen\-era\underline{t}ion)
a commit message generation technique that aims to overcome the above-mentioned limitations of past research.
Specifically, to improve the quality of the training dataset,
we filter out substandard and poor-quality messages.
We propose a new mechanism \deltagraph{}, a graph-based representation of code changes, to capture the context of the source code effectively by generating a union of code, before and after the code changes in each commit.
Our commit message generation module leverages state-of-the-art Encoder-Decoder models fine-tuned on \deltagraph{} results to generate high-quality commit messages.
Finally, we introduce a \textit{quality assurance} module based on Graph Neural Network (\textsc{gnn})
to ensure that our approach selects the best possible message among the generated options based on the format required by the organization.
We evaluate our approach using a variety of performance metrics used by similar studies in the field, including \bleu{}, \rogue{}, and \meteor{}. 
Specifically, we employ \bleu-norm{}~\parencite{loyola2018}, 
a variant of \bleu{} due to its strong correlation with human judgement ~\parencite{tao2021}.
The results of our evaluation indicate that our approach significantly outperforms the state-of-the-art technique in terms of these metrics. 
Specifically, we achieved a \bleu{} score of 7.38, a \rogue{} score of 13.45, and a \meteor{} score of 12.26, which are substantially higher than the state-of-the-art approach. 
This represents an increase of 17.7\% for \bleu{}, and 3\% for \meteor{} compared to the state-of-the-art results, while it remains comparable in terms of \rougel{}.
We also evaluated our approach against GPT models (gpt-3.5-turbo and gpt-4-turbo). We found our approach outperform these models on five and four metrics respectively and provide competitive results with the rest of the metrics.
These results demonstrate the effectiveness of our representation of code changes in generating high-quality commit messages and show promise in applying \deltagraph{} to similar software engineering problems.

Our paper makes the following contributions to the field.
\begin{itemize}
    \item The study proposes \textbf{\deltagraph{}}\textemdash{}a novel graph-based representation of code changes that incorporates the context along with code modifications. 
    Researchers in the field may use and extend \deltagraph{} for software engineering problems where changes in the code with context are required.
    \item The study provides a filtered dataset of high-quality commit messages that can be used in similar studies in the future.
    \item The paper implements a fine-tuned encoder-decoder model trained using a filtered training dataset that achieves state-of-the-art performance in generating commit messages.
    \item The study offers a customizable quality assurance module that enables us to choose the best message among available options where \textit{best} is defined by the employed customizable strategy by the quality assurance module.

\end{itemize}

\noindent
\textbf{Replication package:}
We have made our tool and scripts publicly available 
~\parencite{anonymous_github_2023} to facilitate other researchers to replicate, reproduce, and extend our study. 
We have also provided the filtered dataset of high-quality commit messages online~\parencite{anonymous_2023_models}.

\section{Methods}

This section provides an overview of methods and mechanism to implement our proposed approach.

\subsection{Overview}
The \textit{goal} of the research is to implement a commit message generation approach that considers the context of code changes and produces a high-quality commit message.
We derive the following research questions based on the goal of the presented study.

\begin{description}
    \item [RQ1.] \textbf{Does capturing the code context improve the accuracy of the commit message generation model?}
\end{description}

A code change is not an isolated transformation and therefore, changes must be understood considering the surrounding context.
The proposed research question aims to determine whether capturing the code context improves the accuracy of the commit message generation model. 
Answering this research question will provide insights into the role of the context while generating a commit message.


\begin{description}
\item [RQ2.] \textbf{How does the proposed approach perform compared to existing state-of-the-art automated commit message generation approaches?}
\end{description}

Through this research question, we compare our approach with existing commit message generation methods. Furthermore, we qualitatively evaluate our approach, comparing it against the best performing baseline. 
Exploring this question will show whether and to what extent the proposed technique improve the state of the art.

\begin{description}
\item [RQ3.] \textbf{How does the proposed approach compare against the latest GPT models?}
\end{description}
GPT-3.5 and GPT-4 models, given an appropriate prompt, generate natural language or code. The \texttt{gpt-3.5-turbo} is the most popular model, while \texttt{gpt-4-turbo} is the most capable model from OpenAI at the time of writing this text. 
They are optimized for chat applications; the models currently power ChatGPT~\parencite{chatgpt2022}, a popular chatbot used in various abstractive and extractive tasks belonging to diverse domains~\parencite{dwivedi2023}.
The research question aims to evaluate the performance of the proposed approach against these models to provide insights into the comparison between learning from a large amount of generic data and learning from a small amount of relevant data.



\begin{figure*}[htbp]
    \centering
    \includegraphics[width=\linewidth]{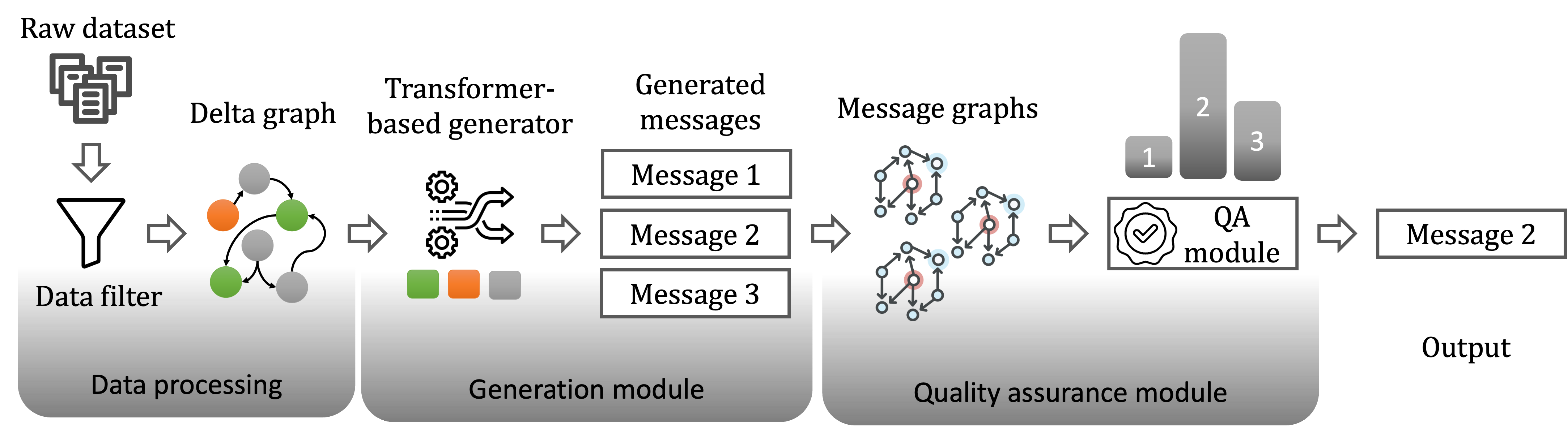}
    \caption{Overview of the approach}
    \label{fig:overview}
\end{figure*}

Figure~\ref{fig:overview} presents an overview of the proposed approach.
The approach uses a pre-processing module that applies a set of filters, to filter out low-quality commit messages from the training dataset. 
Also, we convert code from the filtered dataset into \deltagraph{} representation using our \deltagraph{} module.
This representation is then passed through a transformer-based \dl{} module that learns to generate commit messages given code modifications and its context in the \deltagraph{} format. 
The generator module generates multiple commit messages.
We employ a \textit{Quality Assurance} (QA) module
that ranks messages based on user-defined criteria
to ensure emitting a high-quality message.
Such a mechanism results in a versatile approach that prioritizes relevant and informative messages as per users needs. 
We elaborate on the individual component of our approach in the rest of the section.

\subsection{Data pre-processing}\label{AA}

We train and evaluate \comet{} on the \textsc{mcmd} dataset proposed by~\citet{tao2021}.
We choose this dataset because first,
it has been used in other similar studies~\parencite{shi2022,dey2022}.
Also, the dataset provides corresponding complete code in addition to code diff that is crucial for generating high-quality commit messages.
Also, the dataset offers the ability to trace back each code diff to the original repository, enhancing control over the dataset. 
The dataset originates from the top $100$ starred projects on GitHub, representing a diverse range of code repositories in six different programming languages. 
We limit the scope of our study to the Java programming language. 
Though we train and evaluate our model on a Java dataset, the only code-dependent element in our approach involves parsing the code into \cpg{}. Consequently, our approach can adapt to any programming language, provided a language-dependent tool or library to transform source code into \cpg{}.

After obtaining the dataset consisting of $450,000$ commits, we filter out
files with extensions other than \texttt{.java}. 
Similar to previous approaches in this domain~\cite{jiang2017}, we truncate the commit messages to the first line as it is considered the summary of the whole message. Researchers have found that the initial sentence in commit messages~\parencite{jiang2017} and API comments~\parencite{gu2016deep, javadoc_tool_2019} usually encapsulates the key information and context.
Recognizing the significance of data quality in training deep learning models, we implement measures to use high-quality commit messages in our study.
To achieve the goal, we employ the message characteristic filters proposed by ~\citet{tian2022} that focus on capturing the "why" and "what" aspects of the commit messages. 
The data is passed through two Bi-LSTM layers, which are part of these filters, and obtain their code property graph (\cpg{}) using \texttt{Joern}~\parencite{joern2022}. 
Finally, we obtain $18,214$ commit messages that meet the filter criteria for model training.

\subsection{Delta Graph}\label{sec:DeltaGraph}
An abstract syntax tree (\abst{}) provides a detailed breakdown of source code into language constructs~\parencite{yamaguchi2014} giving a structure-rich representation of the code.
Similarly, a control flow graph (\cfg{}) exhibits the execution flow of the program~\parencite{yamaguchi2014}.
Furthermore, a program dependence graph (\pdg{}) models data and control dependencies of a program.
Previous works on commit message generation remained limited to the use of \abst{} and do not explore alternative representation methods.
Although \abst{}s have long been an efficient code modeling approach,
other information such as data and control dependencies could be leveraged to enrich the representation. 
An information-rich code representation in this context is Code Property Graph (\cpg{}) that combines \abst{}, \cfg{}, and \pdg{}.
This combined representation has been effective in various software engineering tasks such as vulnerability analysis~\parencite{zhou2019,Xia2018} and code refactoring suggestions~\parencite{Cui2023}

\begin{figure}[htbp]
    \centering
    \includegraphics[width=0.7\linewidth]{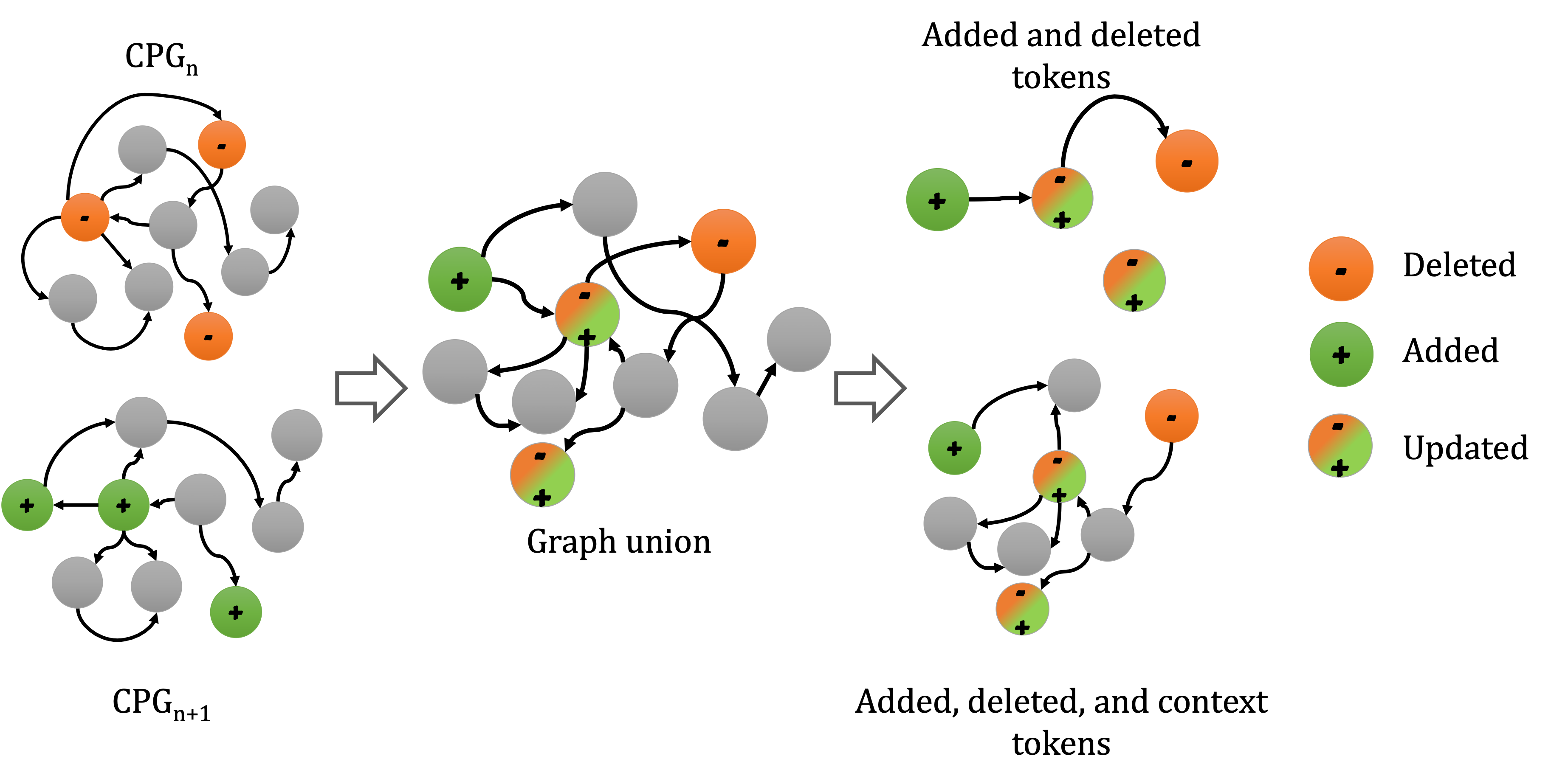}
    \caption{Overview of \deltagraph{} creation}
    \label{fig: DeltaGraph}
\end{figure}

A \deltagraph{} models changes through structural (\ie{} \abst{}) and semantic (\ie{} \cfg{} and \pdg{}) aspects into a single graph representation to maximize the information gained from the changes made.
We utilize \textit{Joern}~\parencite{joern2022}, 
a widely used source-code analysis platform to generate \cpg{}s for code snippets.
We use the generated \cpg{}s, manipulate them, and construct corresponding delta graphs as shown in Figure~\ref{fig: DeltaGraph}.


In order to create a \deltagraph{} that represents the code changes between successive commits,
we transform each version of the program into a \cpg{}. This results in two CPGs denoted as $G_{p1} = (E_{p1}, V_{p1})$ and $G_{p2} = (E_{p2},V_{p2})$, where $G_{p1}$ and $G_{p2}$ correspond to $CPG_{n}$ and $CPG_{n+1}$ respectively as shown in in Figure ~\ref{fig: DeltaGraph}.
Subsequently, we follow the steps given below to build three sub-graphs, aggregate and manipulate them to obtain a \deltagraph{} representation($\Delta G$): 
\begin{itemize}
    \item The common edges and vertices in $G_{p1}$ \& $G_{p2}$ are extracted to represent a common graph $G_{c}$ = ($E_{c}$,$V_{c}$). This sub-graph contains the parts of the code that remain \textit{unchanged}.
    \item The edges and vertices that are present in $G_{p1}$ but not in $G_{p2}$ are combined to form $G_{del}$ = ($E_{del}$,$V_{del}$). This sub-graph contains the \textit{deleted} parts of code. 
    \item The edges and vertices that are present in $G_{p2}$ but not in $G_{p1}$ are considered as \textit{added} tokens. The extraction of these properties results in $G_{add}$ = ($V_{add}$,$E_{add}$). This sub-graph contains code tokens added or, in some cases, updated in the code.
\end{itemize}

These three sub-graphs are then aggregated to form a graph union: $G_{U}$ = $G_{add}$ $\cup$ $G_{del}$ $\cup$ $G_{c}$. 
Graph union is a complete representation of the code change since it considers all possible tokens involved in the code change. 
The core contributors to a commit message are the uncommon tokens between two consecutive commits.
Retaining all tokens leads to redundant information that not only makes the learning expensive but also possibly hurts the performance of the model.
In order to tackle this issue and to extract the relevant contextual tokens that are closely related to the added and deleted tokens, we further process the penultimate representation to only retain the following information: 
\begin{itemize}
    \item Added edges and vertices : $G_{add}$
    \item Deleted edges and vertices : $G_{del}$
    \item Direct edges of added ($E_{add}$) / deleted ($E_{del}$) edges : $G_{[c]}$, where $G_{[c]}$ $\subseteq$ $G_{c}$
\end{itemize}

The decision to restrict the context to direct edges is inspired by the notion of over-smoothing~\cite{li2018} in graph neural networks. 
The objective is to pass the most relevant features as context and as we move away from the modified tokens, the significance diminishes, providing a rationale for this approach. 
Moreover, considering additional edges elongates the context length, 
thereby posing a risk of performance degradation in \textsc{llm}s~\cite{liu2023}.
The retained representations are united to form a \deltagraph{} representation: $\Delta G$ = ($\Delta E$, $\Delta V$), where
$\Delta E$ = $E_{add}$ $\cup$ $E_{del}$ $\cup$ $E_{[c]}$
and 
$\Delta V$ = $V_{add}$ $\cup$ $V_{del}$ $\cup$ $V_{[c]}$.



\subsection{Generation module}\label{Sec:generation}

We leverage transformer~\parencite{Vaswani2017} models for their powerful sequence generation ability to achieve the goals of the study.
The input to transformers is usually a sequence of tokens that are mapped to their corresponding embeddings. 
We aggregate these embeddings with positional embeddings forming an initial vector representation. 
As shown in Equation~\ref{eqn:self-attention}, we pass the initial vector representations through a self-attention layer and a feed-forward neural network. 
The encoder's final representation is then passed onto the decoder through cross-attention layers.
The decoder follows a similar architecture as the encoder except for an additional layer with cross-attention input from the encoder. 
The decoder outputs the sequence regressively using masked-self attention as shown in Equation~\ref{eqn:self-attention}, ensuring it can only attend to the past tokens to generate a new word in the sequence.

\begin{equation}
\label{eqn:self-attention}
\text{SelfAttention}(Q,K,V) = \text{softmax}(\frac{QK^T}{\sqrt{d_k}})V
\end{equation}

The multi-headed self-attention block comprises several attention heads that operate in parallel and are concatenated.

\begin{equation}
\textrm{MultiHead}(Q, K, V) = \textrm{Concat}(\textrm{head}_1, \ldots, \textrm{head}_h) W_O
\end{equation}

\noindent
where
\begin{equation}
\textrm{head}_i = \textrm{Attention}(QW_i^Q, KW_i^K, VW_i^V)
\end{equation}

Pre-trained transformers have shown huge success across various fields due to their dominant capability in zero-shot or few-shot generalization~\parencite{zhang2022opt}. 
Pre-trained transformer models have proven effective in various code-related tasks such as code summarization, code comment generation, and others~\parencite{Sharma2024}. 
Several studies on commit message generation~\parencite{shi2022,dong2022,liu2020} leveraged encoder-decoder-based architecture to generate commit messages and have achieved promising results proving to be effective in this domain. 

Commit message generation is a \textit{code-to-text} task that requires a strong understanding of both code and text. 
Transformer-based encoder-decoder models such as CodeT5~\parencite{wang2021codet5}, 
and auto-regressive models in recent times such as GPT-3~\parencite{brown2020} and BLOOM~\parencite{scao2022}, 
have demonstrated their ability in sequence-to-sequence generation tasks in both natural language and code. 
Fine-tuning is a common approach used to leverage the transformer's domain knowledge. 
Pre-trained models require minimal data to adapt to a given task and can thus significantly reduce the time and resources required to train a model specific to a downstream task.

\comet{} fine-tunes a pre-trained transformer on \deltagraph{} representation. 
As the transformer's encoder or decoder accepts a linear sequence of tokens as input, the \deltagraph{} representation is linearized in a sequence of added, deleted, and common tokens
and passed into the transformer. The delta graph captures structural and semantic aspects of code changes, including added, deleted, and context tokens, which helps retain important information about the code. By leveraging the delta graph representation, the model effectively capture and utilize the structural information of the code changes.


There are several pre-trained transformers trained on code that are publicly available for use such as BLOOM~\cite{scao2022}, CodeT5~\cite{wang2021codet5}, GraphCodeBERT~\cite{guo2021}, and CodeBERT~\cite{feng2020}.
We select a set of pre-trained transformer-based models to experiment with, where either the encoder, or the decoder, or both are trained on a combination of code and natural language.
A combination of GraphCodeBERT~\cite{guo2021} \& GPT-2~\cite{radford2019language}, Code-T5~\cite{wang2021codet5} and BLOOM~\cite{scao2022} are the models satisfying the above criterion.
The selection exercise is important as we intend to evaluate the selected pre-trained models to identify the best-performing model for \comet{}.
We evaluate the selected models against three performance metrics: \bleu{}, 
\rougel{}, 
and \meteor{}.

\subsection{Quality assurance module}
\begin{figure*}[ht]
    \centering
    \includegraphics[width=1\linewidth]{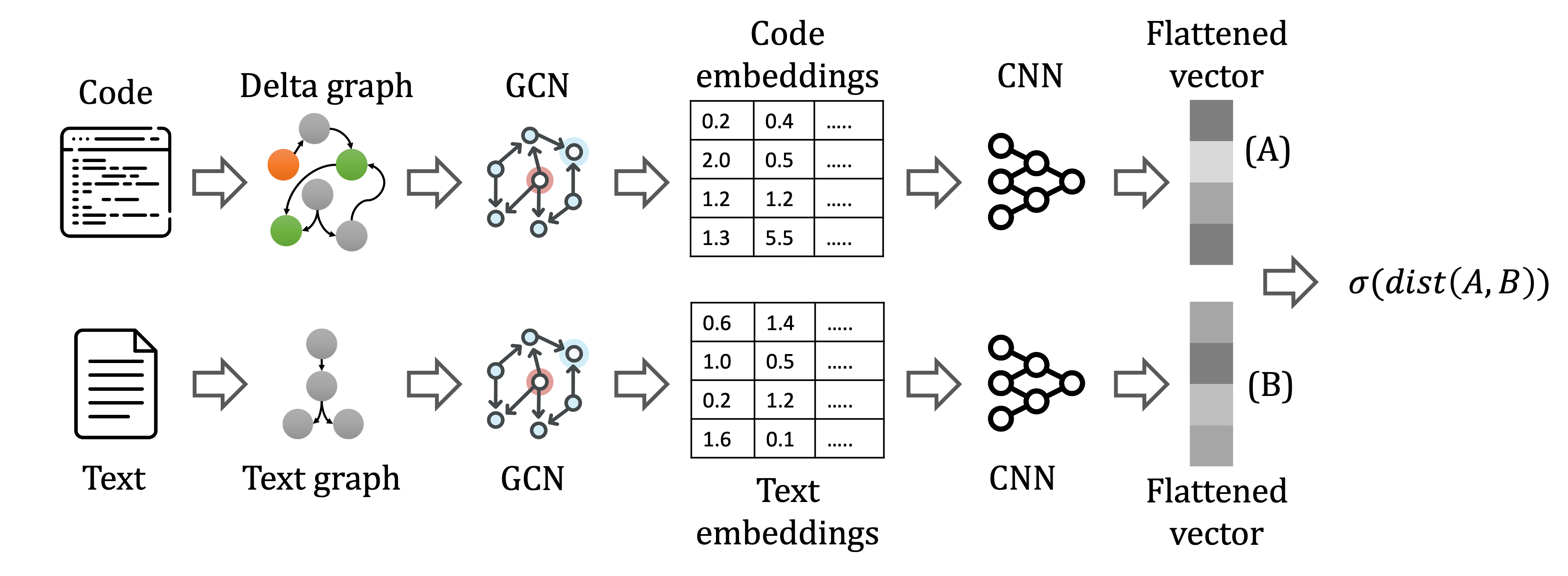}
    \caption{Architecture of Quality Assurance module}
    \label{fig:architecture-qa}
\end{figure*}

To tackle organization or developer specific subjectivity in commit message generation, 
\comet{} proposes a Quality Assurance (QA) module that ensures the selected messages meet the user-specific needs and adhere to their criteria. 
To assess the generated messages from the Generation module, 
the QA module incorporates rich embeddings for both code and text, prioritizing the most relevant and informative messages for the user. 
In addition, the module leverages a graph-based representation of code and text, 
formed from a Graph Convolutional Network (\textsc{gcn})~\parencite{kipf2016}
and passed onto a paired Convolutional Neural Network (\cnn{}) for training. 
This approach allows, depending upon the training data for QA module,
a customized organization-specific commit message selection among the messages generated by the Generator module.
Below, we elaborate on the steps involved in implementing the QA module.

\begin{itemize}
    \item 
    After acquiring the \deltagraph{} as described in section 
    \ref{sec:DeltaGraph},
    we pass code tokens corresponding to the edges of \deltagraph{} ($\Delta G$) into CodeBERT~\parencite{feng2020} and extract the \texttt{[CLS]} token embedding to act as initial representation for Edges ($\Delta E$). 
    The \deltagraph{} is then passed into a Graph Convolution Network trained on node classification task for predicting the type of Edge (added ($E_{a}$), deleted ($E_{d}$) or common ($E_{c}$) to obtain the final representations.
    \item Words from a commit message are parsed into a dependency parse tree using the StanfordNLP parser~\cite{manning2014} to leverage the grammar dependency property between words in a sentence. 
    \item Word tokens corresponding to the edges of the dependency parse tree ($G_{text}$) are passed into BERT, and the \texttt{[CLS]} embeddings are extracted to serve as the initial edge ($E_{text}$) representations. 
    The text graph is passed into a Graph Convolution Network trained on an incoming vertex ($V_{text}$) class prediction task to produce a final representation of the text graph.
    \item The final graph node representations are aggregated and are of size \texttt{[len($E_{\Delta/text}$)$\times$768]}, where \texttt{len($E_{\Delta/text}$)} is the number of Edges in the graph.
\end{itemize}

To train the QA module, we use a paired Convolutional Neural Network with shared weights. 
The training process involves labeling the \textit{delta graph - message} pairs in binary notation based on user preferences for message quality (1: preferred and 0: not preferred). 
This approach treats the training as a binary classification problem. 
The paired CNN takes the aggregated representations from both the text and code Graph Neural Networks (GNNs) as input. 
The penultimate layer of the paired CNN is flattened and is passed through an additional neural network layer to obtain the final vector representations of code and text. 
We then calculate the Euclidean distance between these vector representations and pass the resulting scalar through a sigmoid activation as given in Equation~\ref{eq:sigmoid} to generate the probability of the code-message pair belonging to the positive class.

\begin{equation}
\label{eq:sigmoid}
    \begin{gathered}
        \mathbf{\sigma(x) = \frac{1}{1 + e^{-x}}}
    \end{gathered}
\end{equation}
\[
where,\ x = distance(\mathbf{p}, \mathbf{q}) = \sqrt{\sum_{i=1}^{512} (q_i - p_i)^2}
\]

The QA module receives multiple code-message pairs at the inference stage and calculates their probabilities. 
These probabilities are then used to rank the messages in descending order of their relevancy to the code change. 
The ranking of commit messages depends on the training samples and the chosen criteria. 
The current version of \comet{} maintains a uniformity between the results generated by the decoder and the QA module's preference ranking as it is being offered as a flexible module to be moulded by the user itself. However, in the future, we plan to explore and implement different QA strategies that are tailored to the requirements of different organizations.

\subsection{Experimental setup}
After experimenting with various learning rates such as $5\times10^{-5}$, $1\times10^{-5}$, and $2\times10^{-5}$, we found that a learning rate of $5\times10^{-5}$ worked best for our model. This learning rate is also the default learning rate for the Adam optimizer, as shown in the official Hugging Face documentation~\parencite{learningrate2022}. We implemented and trained the model using the Hugging Face implementation of transformers~\parencite{wolf2020}, and used AdamW~\parencite{loshchilov2017} optimizer with a learning rate of $5\times10^{-5}$ for fine-tuning the model over a maximum of $9$ epochs, with a batch size of 16. 
We set the maximum input length to $512$, and the maximum generated message length to $80$, based on the data dimensions and the model's input limit. The models were trained on both a single \textsc{nvidia} P100 \textsc{gpu} and a single \textsc{nvidia} V100 \textsc{gpu}.

\subsection{Metrics}
We evaluate \comet{} on three popular metrics: \textsc{bleu}~\parencite{papineni2002}, \rougel{}~\parencite{lin2004}, and \meteor{}~\parencite{banerjee2005} to demonstrate its efficiency. 
\textsc{bleu} is a widely used precision-based metric that calculates the n-gram overlap between the predicted and reference candidates and averages them over the whole corpus. All the  \textsc{bleu} results reported in this paper are of \textsc{bleu-norm}~\parencite{loyola2018} variant.
We evaluate and report up to 4 n-grams for \bleu{}.
\rogue{} is a recall-based metric that measures the overlap between the predicted and reference candidates in terms of n-gram matches. We use the version of \rogue{} that calculates the score based on the longest common subsequence (\textsc{lcs}) between the predicted and reference candidates.
\meteor{} is an F1-based metric that measures the quality of the generated text by mapping contiguous predicted unigrams to contiguous unigrams in the references and computing their harmonic mean with respect to precision and recall.
These metrics have been commonly used in existing commit message generation techniques~\parencite{liu2020,shi2022,jung2021}.
\section{Results}
In this section, we present our observations from the experiments for the considered research questions.


\subsection{Results of RQ1}

In RQ1, we aim to evaluate the role of context tokens 
in boosting commit message generation accuracy.
\comet{} uses \deltagraph{} to capture the context along with the code changes.
For the experiment, we choose the following three pre-trained models that we selected based on the criteria mentioned in Section~\ref{Sec:generation}.

\begin{enumerate}
    \item 
    \textbf{GCB-GPT2:} 
    The first model is formed using GraphCodeBERT~\parencite{guo2021} as the encoder and GPT-2~\parencite{radford2019language} as the decoder.
    GraphCodeBERT is pre-trained on code with pre-training tasks such as edge prediction in data-flow graphs, equipping it with powerful programming knowledge. 
    GPT-2 is a popular open-source auto-regressive model trained on text data 
    (English corpus majorly) and works well in text generation tasks favouring commit message generation.
    \item 
    \textbf{CodeT5:} 
    It is a powerful encoder-decoder model with the architecture borrowed from T5 transformers. 
    CodeT5's challenging pre-training tasks make it a robust model that has demonstrated its effectiveness across various code-related tasks~\parencite{wang2021codet5}.
    \item 
    \textbf{BLOOM:}
    The BLOOM model~\parencite{workshop2023bloom} has shown promising results across various zero-shot generalization tasks in code and natural language.
\end{enumerate}

\begin{table}[ht]
    \centering
    \caption{RQ1: Comparison of models' performance}
    \label{tab:rq1-comparison}
    \adjustbox{max width=\columnwidth}{
    \begin{tabular}{@{}llllllll@{}}
    \toprule
    \textbf{Model}            & \textbf{Data}         & \textbf{BLEU 1} & \textbf{BLEU 2} & \textbf{BLEU 3} & \textbf{BLEU 4} & \textbf{METEOR} & \textbf{ROGUE-L} \\ \midrule
    GCB-GPT2       & ADC          & 10.09  & 3.9    & 1.9    & 1.0      & 10.3      & 8.2\\
                     & AD           & 9.6    & 3.7    & 1.76   & 0.91   & 9.84   & 7.76\\[1ex] 
    \textbf{CodeT5} & \textbf{ADC} & \textbf{17.33} & \textbf{8.96} & \textbf{5.4} & \textbf{3.53} & \textbf{12.26} & \textbf{13.45}\\
                     & AD           & 15.58 & 7.83 & 4.73 & 3.11 & 11.49 & 11.85\\[1ex]
    BLOOM            & ADC          & 6.12   & 1.9    & 0.8    & 0.4    & 5.49   & 5.9\\
                     & AD           & 8.85   & 3.1    & 1.3    & 0.6    & 6.87   & 6.88\\ \bottomrule
    \end{tabular}
    }
\end{table}

We fine-tune the models with the following two types of data:   
\textbf{AD} (containing added and deleted tokens)
and
\textbf{ADC} (containing added, deleted, and context tokens).
Table~\ref{tab:rq1-comparison} presents results from the experiment.
The findings demonstrate that CodeT5 outperforms the other two models across all the metrics in both the data representations (AD and ADC) supplied as input. 
We have also observed that context tokens facilitate in $7\%$ and $13.4\%$ boost in performance on the BLEU metric for GCB-GPT2 and CodeT5  respectively. 

Although context tokens demonstrate a performance boost over all the metrics for the first two models, an interesting finding is that BLOOM, a decoder-only model shows an inferior performance with context tokens. One potential reason for the inferior performance of BLOOM with context tokens could be that the decoder-only architecture of BLOOM is not well-suited for capturing and incorporating context information. Unlike encoder-decoder models that have a separate component for encoding the input context, BLOOM solely relies on the decoder to generate the output tokens, which might not effectively utilize the contextual information.
Further detailed exploration and experimentation is required to explain the performance degradation for the generative model. 
Even though BLOOM without context did better than with context, its performance is significantly lower than the other two models.

\begin{rqbox}
    \textbf{Summary of RQ1:}
    Our findings suggest that incorporating contextual representation significantly improves the performance of encoder-decoder based models.
    We observe an improvement of $7\%$ and $13.4\%$ on the BLEU metric for considered Encoder-Decoder models.
\end{rqbox}


\subsection{Results of RQ2}
In this research question, we evaluate our model's performance against the baseline, including the existing state-of-the-art approaches.
For this comparison, we identify the most relevant and latest commit message generation techniques \ie{}
RACE~\parencite{shi2022}, NNGen~\parencite{Liu2018}, ATOM~\parencite{liu2020}, and Commit-BERT~\parencite{jung2021}.
RACE is a hybrid approach for commit message generation that utilizes a generation and retrieval module.
NNGen is an information-retrieval-based model that retrieves the top-k commits and returns the one with the highest 4-gram precision match.
ATOM utilizes \abst{}s to represent code differences and leverages a \textit{tree2seq} model to generate commit messages.
However, we could not find their replication package;
we reached out to the authors of the study and did not receive any response.
We also considered FIRA~\parencite{dong2022} as one of our baseline approaches;
however, we could not reproduce its implementation. Specifically, multiple implementation issues in their pre-processing scripts hindered our plans to include FIRA in our comparison.
 

Commit-BERT utilizes a pre-trained model CodeBERT to generate commit messages. 
The work first fine-tunes the model on CodeSearchNet~\parencite{husain2019} dataset to inject the domain knowledge into the model and then trains the model on the commit message generation dataset.
We carry out our experiment on the selected baseline models along with our proposed approach \comet{}.
We train the models on our dataset and document the reported performance metrics.
We present the results of the experiment in Table~\ref{tab:rq2}.

\begin{table}[ht]
\caption{Comparison of \comet{} with baseline approaches}
\label{tab:rq2}
\adjustbox{max width=\columnwidth}{
\begin{tabular}{@{}llllllll@{}}
\toprule
    \textbf{Baseline}      & \textbf{BLEU 1} & \textbf{BLEU 2} & \textbf{BLEU 3} & \textbf{BLEU 4} & \textbf{METEOR} & \textbf{ROGUE-L} \\ \midrule
RACE            & 13.5  & 7.3  & 4.7 & 3.3  & 11.9  & \textbf{13.6 } \\
Commit-BERT   & 1.70    & 0.70     &  0.30   &  0.20    &  1.74     & 1.20      \\
NNGen         & 3.5     & 1.6     &1.02   &0.7  &  6.03  &  6.21      \\
\textbf{COMET}          & \textbf{17.33} & \textbf{8.96} & \textbf{5.4} & \textbf{3.53} & \textbf{12.26} & 13.45 \\ \hline
\end{tabular}
}
\end{table}

The results indicate that \comet{} outperforms all the evaluated approaches in all metrics except for \textsc{rogue-l}, where ATOM perform better than \comet{}.
Our model improves $17.7\%$ on \bleu{} metrics compared to the existing state-of-the-art approach. We also provide an example of generated commit messages for code diff presented in Listing~\ref{fig:commit-messages} to gain more insight into the performance of the evaluated approaches.
It shows that though commit message generated by \comet{} is not matching perfectly with the ground truth, \comet{} captures the context of the code change successfully and provides a clearer explanation of the intended changes better than even the ground truth. 

\begin{center}

\begin{listing}[ht]
\begin{minted}[breaklines]{bash}  
diff --git a/LockGraphManager_old.java b/LockGraphManager_new.java
index 24c9111..a5835c5 100644
--- a/LockGraphManager_old.java
+++ b/LockGraphManager_new.java
@@ -174,7 +174,7 @@
\end{minted}
\begin{minted}[breaklines,highlightcolor=red!30,highlightlines={3}]{java}
public abstract class LockGraphManager<LOCK_TYPE extends DBAServerLock<?>, ID_TY
for(LOCK_TYPE root : roots) {
- if (root.waitThis().size() >= 0)
\end{minted}
\begin{minted}[breaklines,highlightcolor=green!30,highlightlines={1}]{java}
+ if (root.waitThis().size() > 0)
 createGraph(root);
 }
\end{minted}


\caption{Sample code change for commit message generation using baseline techniques}
\label{lst:rq2_example}
\end{listing}


{\bf Commit messages:}
\begin{itemize}
\item {\bf Ground Truth:} "fixed small issue with if-condition"
\item {\bf RACE:} "Lock graph manager fix"
\item {\bf NNGen:} "\#315 Pure comment queries fix"
\item {\bf CommitBert:} "[hotfix][tests] Remove unused import"
\item {\bf COMET:} "LockGraph: only create root node if there are any locks."
\end{itemize}
\label{fig:commit-messages}
\end{center}

Motivated by the above example and aiming to avoid demonstrating only the example where our method is giving best results, we conducted a qualitative assessment of our proposed approach in comparison to RACE~\parencite{shi2022}.
RACE is the state-of-the-art model for commit message generation. 
To evaluate and compare the performance of both the approaches,
we created a survey inspired by similar studies~\parencite{liu2020, dong2022}. 
We randomly sampled $50$ data points from the test set for the evaluation process.
Every data point includes a reference message, along with two predicted messages generated separately by RACE and \comet{}. 
We invited graduate student volunteers to participate in this assessment from the Faculty of Computer Science of Dalhousie University.
Three PhD students with prior software development experience volunteered
to assess the predicted messages against the ground truth data. 
To minimize bias, we ensured that the model responsible for each predicted message remained anonymous to the participants. 
The participants were provided with an existing scoring criteria from previous similar studies~\parencite{Liu2018, dong2022} as shown in Table~\ref{table:scoring_information}.

\begin{table}[ht]
\caption{Commit message scoring criteria for manual assessment}
\label{table:scoring_information}
\adjustbox{max width=\columnwidth}{
\begin{tabular}{ll}
\toprule
Score                   & Description    \\ \midrule
0 & Neither relevant in semantic nor having shared tokens \\
1 & Irrelevant in semantic but with some shared tokens  \\
2 & Partially similar in semantic, but contains exclusive information \\
3 & Highly similar but not semantically identical  \\
4 & Identical in semantic \\ \bottomrule
\end{tabular}}
\end{table}

\begin{figure}[ht]
    \centering
    \includegraphics[width=0.6\linewidth]{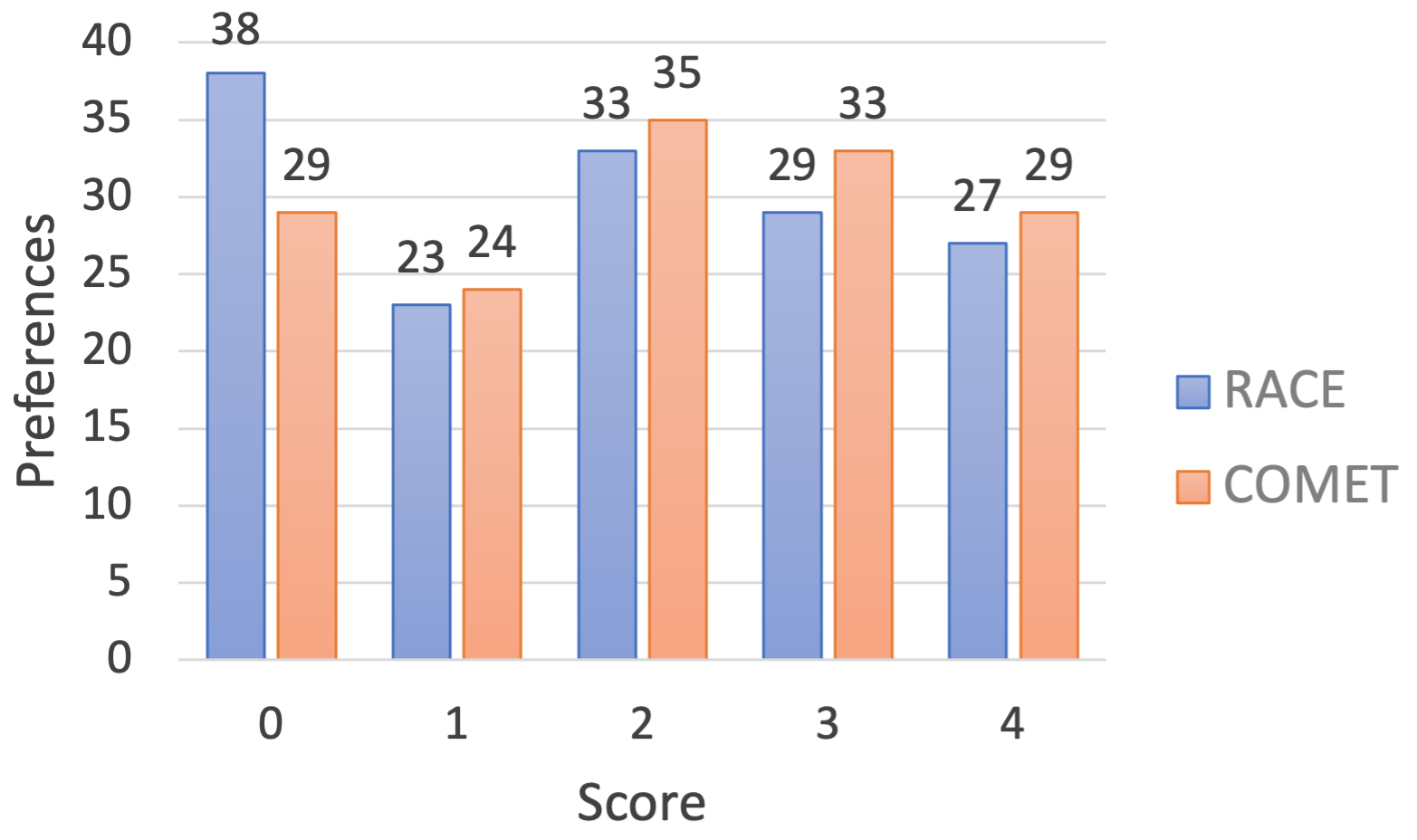 }
    \caption{Results of qualitative assessment}
    \label{fig:qualitative-study}
\end{figure}

Each participant assessed all the $50$ samples and assigned a score. Hence, we obtained a total of $150$ observations.
Figure~\ref{fig:qualitative-study} presents the distribution of all the obtained observations for RACE and \comet{}. 
RACE messages scored an average of $2.56$, while \comet{} achieved an average of $2.72$.
\textbf{Therefore, we conclude that COMET performed better than RACE in an anonymous qualitative assessment.}

Table~\ref{table:qa_samples} shows
three samples randomly picked from the evaluation set.
One may find their corresponding code diffs in our replication package~\parencite{anonymous_github_2023} in \textit{Qualitative-Evaluation} folder.
These examples show that \comet{} generates commit messages with more contextual information compared to the state-of-the-art approach or even the reference message.

\begin{table}[ht]
\caption{Qualitative Assessment samples}
\label{table:qa_samples}
\adjustbox{max width=\columnwidth}{
\begin{tabular}{ll}
\toprule
Reference-1 & RateLimiter fix (\#2229)  \\ 
RACE        & refactoring \\
COMET       & Update RateLimiterLongRunningUnitTest.java  \\
\toprule
Reference-2 & refactor demo\\
RACE        & refactoring  \\
COMET       & apollo config demo improvements \\ 
\toprule
Reference-3 & rename to assertGenerateKeyWithClockCallBack()\\
RACE        & Update DefaultKeyGeneratorTest.java  \\
COMET       & Change test method name to use new method name \\ 
\bottomrule
\end{tabular}}
\end{table}

\begin{rqbox}
    \textbf{Summary of RQ2:}
    The results show that \comet{} surpasses the state-of-the-art model by an average of $12.2\%$ across all six metrics, establishing itself as an effective approach for generating high-quality commit messages. Furthermore, our manual comparative assessment shows a stronger preference for \comet{} compared to the state-of-the-art approach.
\end{rqbox}

\subsection{Results of RQ3}
ChatGPT~\parencite{chatgpt2022}\textemdash{}a transformer-based model optimized for chat trained using Reinforcement Learning with Human Feedback (RLHF), has shown remarkable capability in performing code and text-related tasks~\parencite{laskar2023}.
Therefore, in this research question, we aim to compare the performance of our proposed approach against the \textit{gpt-3.5-turbo} and \textit{gpt-4-turbo} models that power ChatGPT~\parencite{chatgpt2022} in the commit message generation task. We conduct an extensive evaluation and present findings across zero-shot, one-shot, and multi-shot settings.

Prompt engineering has emerged as a new field of interest with the rise of \textsc{llm}-powered chatbots. 
Previous research has identified various parameters that define a good prompt for \textsc{llm}-based software such as \chatgpt{}, including factors such as prompt length, specificity, diversity, and relevance criteria~\cite{white2023prompt,tian2023}. 
Following the recommended practices in existing research~\cite{white2023prompt,tian2023}, 
we identified three key criteria for generating a prompt.

\begin{itemize}
    \item \textbf{Completeness}: 
    Prompts should clearly and concisely describe the task at hand.
    \item \textbf{Relevance}: 
    Prompts should not contain any misleading or extraneous information.
    \item \textbf{Comprehensiveness}: 
    Prompts should be sufficiently detailed to provide the necessary information.
\end{itemize}

After carefully considering the aforementioned criteria along with GPT best practices suggested by OpenAI~\parencite{GPTbestpractices}, we generated five diverse prompts that can be found in our replication package\footnote{\url{https://github.com/SMART-Dal/Comet/blob/main/ChatGPT/README.md}}.
To keep the prompt selection process unbiased, we conducted a survey. We selected a diverse set of participants with software development experience to evaluate the effectiveness of the prompts. 
The participants were three Ph.D. students, four master's students, and one bachelor's student.
We provided them with the code diff, reference message, and the message predicted by gpt-3.5-turbo for all three settings (\ie{} zero-shot, one-shot, and multi-shot) mentioned. The participants were asked to choose the top three prompts in each scenario following the criteria defined above, wherein the predicted messages are most similar to the reference message.

\begin{figure}[ht]
    \centering
    \includegraphics[width=0.7\linewidth]{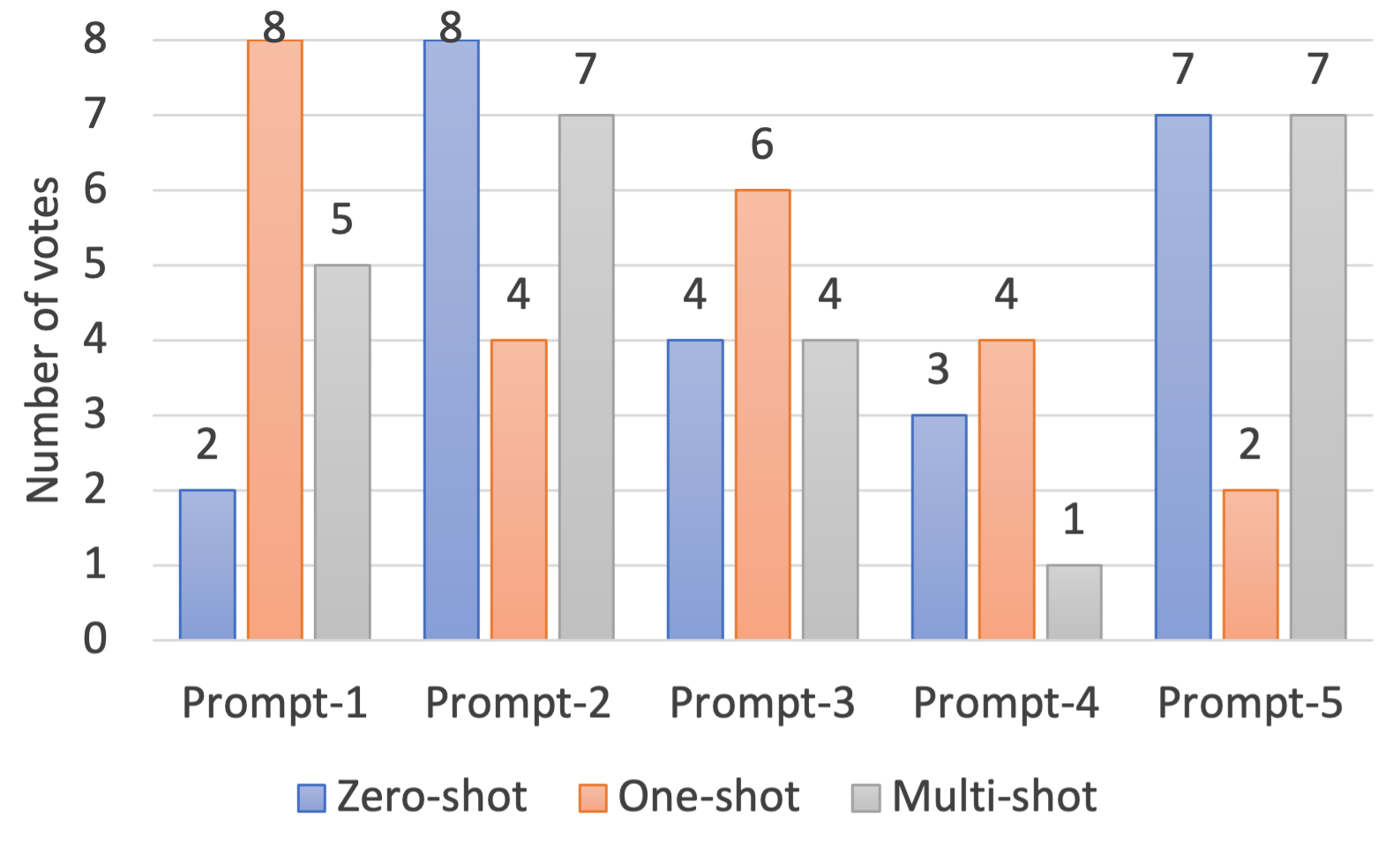}
    \caption{Prompt selection survey}
    \label{fig:prompt-selection}
\end{figure}

Figure \ref{fig:prompt-selection} shows the votes received for each prompt across previously mentioned cases from the participants. By using voting, we obtained a consensus view of the best prompts to use for each case in our evaluation study.
It also helped ensure that the prompts were high quality and met the criteria for this study.
The survey results, selected prompts and messages generated by gpt-3.5-turbo and gpt-4-turbo can be accessed online\footnote{\url{https://github.com/SMART-Dal/Comet/tree/main/ChatGPT/RQ3}}.



For evaluation, we append code diffs from each commit in the test set to the selected prompts across three scenarios.
Specifically, for one-shot and multi-shot settings, CodeBERT~\parencite{feng2020} is utilized to acquire vector representations of all the code diffs and cosine similarity is employed to identify the most similar code diffs, which serve as examples in this case.
In order to utilize the model to the best of it's capability, we implement the chain-of-thought prompting~\parencite{wei2022} technique. Instead of merely stacking code diffs and messages to be passed as input, we include intermediate reasoning steps that provide a description of the input and expected output. We used the gpt-3.5-turbo and  gpt-4-turbo \api{}s provided by OpenAI to send each of these samples as individual \api{} requests and collect the response for each of them. Once we had all the commit messages, we evaluated them using three metrics: \textsc{bleu-norm}~\parencite{loyola2018}, \rougel{}~\parencite{lin2004}, and \meteor{}~\parencite{banerjee2005}.
We present the findings from our experiments in Table~\ref{tab:chatgpt}.

\begin{table}[ht]
\caption{GPT-3.5-turbo Comparison Results} 
\label{tab:chatgpt}
\adjustbox{max width=\columnwidth}{
\begin{tabular}{llllllll}
\toprule
\textbf{Prompts}    & \textbf{BLEU-1} & \textbf{BLEU-2} & \textbf{BLEU-3} & \textbf{BLEU-4} & \textbf{METEOR} & \textbf{ROUGE-L} \\ \midrule 
Zero-shot P1  & 6.49  & 2.69 & 1.39 & 0    & 11.01 & 9.48  \\
Zero-shot P2  & 10.21 & 4.2  & 2.17 & 1.25 & \textbf{12.3}  & 10.24 \\
Zero-shot P3  & 8.67  & 3.7  & 1.93 & 0.11 & 10.22 & 11.8  \\
One-shot P1   & 12.64 & 6.02 & 3.66 & 2.55 & 11.1  & 11.46 \\
One-shot P2   & 11.11 & 4.88 & 2.67 & 1.67 & 10.92 & 11.3  \\
One-shot P3   & 11.2  & 5.03 & 2.82 & 1.8  & 11.36 & 10.8  \\
Multi-shot P1 & 8.68  & 4    & 2.28 & 1.47 & 10.95 & 11.6  \\
Multi-shot P2 & 4.99  & 2.18 & 1.2  & 0    & 11.42 & 10.5  \\
Multi-shot P3 & 6.71  & 2.9  & 1.57 & 0.96 & 11.27 & 11.44 \\
\textbf{\comet{}}   & \textbf{17.33} & \textbf{8.96} & \textbf{5.4}  & \textbf{3.53} & 12.26 & \textbf{13.45} \\ \hline
\end{tabular}
}
\end{table}

\begin{table}[ht]
\caption{GPT-4-turbo Comparison Results} 
\label{tab:chatgpt4}
\adjustbox{max width=\columnwidth}{
\begin{tabular}{llllllll}
\toprule
\textbf{Prompts}    & \textbf{BLEU-1} & \textbf{BLEU-2} & \textbf{BLEU-3} & \textbf{BLEU-4} & \textbf{METEOR} & \textbf{ROUGE-L} \\ \midrule 
Zero-shot P1  & 4.41  & 1.81 & 0.92 &  0.54  & 12.43 &  6.97 \\
Zero-shot P2  & 9.75 & 3.84  & 1.98 & 1.15 & \textbf{14.58}  &  11.31\\
Zero-shot P3  & 10.10 &  4.28 & 2.21 & 1.29 & 11.11 &  13.17 \\
One-shot P1   & 8.52 & 3.48 & 1.79 & 1.04 & 11.75 & 11.36 \\
One-shot P2   & 7.98 & 2.98 & 1.49 & 0.87 & 13.46 &  10.17 \\
One-shot P3   & 4.74  & 1.86 & 0.91 & 0.51 & 12.31 & 7.13 \\
Multi-shot P1 & 9.53  & 4.50 & 2.66 & 1.77 & 12.90 & 12.48 \\
Multi-shot P2 & 2.54  & 1.08 & 0.60 & 0.39 & 10.48 & 5.33 \\
Multi-shot P3 & 14.26 & 7.37 & 4.70 & 3.38 & 13.31 & \textbf{14.63} \\
\textbf{\comet{}}   & \textbf{17.33} & \textbf{8.96} & \textbf{5.4}  & \textbf{3.53} & 12.26 & 13.45 \\ \hline
\end{tabular}
}
\end{table}
\textbf{Our results in Table~\ref{tab:chatgpt} and Table~\ref{tab:chatgpt4} clearly show that COMET outperform gpt-3.5-turbo model across five metrics and gpt-4-turbo model across four metrics by a significant margin.}
The performance of our approach is comparable with the gpt-3.5-turbo model's zero-shot prompt-2 setting and gpt-4-turbo model's zero-shot prompt-2 and multi-shot prompt-3 settings.
Given that \comet{} uses CodeT5, which has 770 million parameters, while the ChatGPT models using gpt-4 have hundreds of billions of parameters~\parencite{gpt4_wiki}, the performance gain is significant.

\begin{rqbox}
    \textbf{Summary of RQ3:}
    Based on our extensive evaluation, \comet{} demonstrates a competitive capability in generating high-quality commit messages compared to the most popular GPT 3.5 model and the most capable GPT 4 model.
\end{rqbox}
\subsection{Discussion}

The use of structured representations such as \abst{}s and \cpg{}s has shown promise in improving commit message generation. However, the scalability of this approach remains a concern due to the significant computational cost associated with building and manipulating trees or graphs.
Existing studies tackle this issue by truncating the input, which results in loss of information.
This issue calls for advanced techniques to capture relevant features and characteristics without bloating input representations.

Although there are several datasets available, the lack of high-quality commit message datasets is still a significant concern.
Generating high-quality commit messages requires high-quality training data. 
The subpar quality of the ground truth dataset for commit messages can result in underestimating the quality of the generated commit messages even if they are of high quality, leading to unreliable evaluation metrics. Therefore, it is crucial to improve the quality of the ground truth dataset for commit messages for ensuring more accurate evaluation metrics and for achieving higher quality generated commit messages.

Also, current accuracy metrics for evaluating generated commit messages have shown limitations in capturing the message's quality, as discussed in several recent studies, including Dey \etal{}~\cite{dey2022}. 
Construct validity of the commonly used metrics often attract criticism. Therefore, more research is necessary to develop more effective metrics for evaluating the quality of generated commit messages.

\section{Related Work}
Studies on generating commit messages automatically can be broadly classified into four categories: \textit{machine learning-based}, \textit{rule or template-based},\textit{information retrieval-based}, and \textit{hybrid} approaches~\parencite{tao2021}.

\noindent
\textbf{Machine Learning-based Methods:}
The initial work in this domain was limited to classifying changes made to the source code using traditional machine learning techniques.
Such efforts include classifying source code changes to categories of maintenance tasks~\parencite{Hindle2009,levin2017}, and
clustering commits to understand the intent of implementation~\parencite{Yamauchi2014}.
Recent advancements in generation techniques, 
such as neural machine translation and attention mechanisms, 
have shown promising results in generating concise and meaningful commit messages by considering code change information. 
However, early studies, \eg{} neural machine translation (\textsc{nmt})-based approaches~\parencite{jiang2017,loyola2017,vanhal2019} treat code as a flat sequence of tokens,
ignoring syntactic and semantic code change information. Such approaches suffer from a limited vocabulary of the most frequent words and fail to learn the semantics of the code changes.

To overcome these limitations, \textit{CoDiSum}~\parencite{Xu2019} utilizes an attention mechanism and a multi-layer bidirectional \textsc{gru} to consider both code structure and semantic information. 
Additionally, they employ a "copying" mechanism to address the out-of-vocabulary (\textsc{oov}) problem by directly copying words from code diffs to the commit message. 
However, \textit{CoDiSum} falls short in capturing the actual code structure as it treats code as a linear sequence of tokens. 
It also has a limited ability to capture long-term dependencies.
\textsc{Fira}~\cite{dong2022} is a graph-based model that uses fine-grained graphs to represent code changes, a graph neural network in the encoder to encode graph-structured inputs, and a transformer and dual copy mechanism to generate commit messages.

One common and major shortcoming of all of these early deep learning-based approaches is the need for their models to be trained from scratch.
Due to this, these approaches are inefficient and exhibit a low capacity to capture the context between natural language (\textsc{nl}) and programming language (\textsc{pl}). 
This led to the development of large language model-based approaches that leverage pre-trained models trained on large corpora of \textsc{nl}-\textsc{pl} pairs.
For example, CommitBERT~\parencite{jung2021} leverages CodeBERT~\parencite{feng2020}, a code-based language model pre-trained on source code.

\noindent
\textbf{Rule-Based Approaches:}
Buse \etal{}~\parencite{Buse2010} proposed \textit{DeltaDoc}, an automatic technique that generates textual descriptions of source code modifications using symbolic execution and template-based summarization techniques. 
Similarly, Vásquez \etal{}~\cite{linares2015} proposed \textit{ChangeScribe}, a technique to generate natural language commit messages by taking into account various factors, such as commit stereotypes, types of code changes (\eg{} file renames, property updates), and the impact of underlying code changes 
(\ie{} the relative number of methods impacted by a class in the commit). 
Another rule-based approach, \textit{ARENA}, was proposed by Moreno \etal{}~\cite{Moreno2017} to address the problem of generating release notes, which involves summarizing changes made in the source code. \textit{ARENA} extracts information about code changes, such as the files, classes, methods, and variables modified, as well as any deprecated classes, methods, or variables. It then uses predefined templates for the kind of artifact and kind of change to summarize code changes.

Despite the usefulness of these rule-based approaches, they have limited capability in capturing the intent of code changes, as they are only suitable for certain types of code changes. Additionally, they tend to generate verbose messages that lack specificity, making them unsuitable as commit messages without human intervention.

\noindent
\textbf{Information Retrieval-Based Approaches:}
Rastkar \etal{}~\cite{Rastkar2013} proposed the use of multi-document summarization and information retrieval techniques to generate a concise natural language description of code changes. The approach captures the rationale of code changes, providing developers with better information about changes in commits.
Liu \etal{}~\cite{Liu2018} proposed \textit{NNGen}, which relies on a bag-of-words model that calculates cosine similarity between the target code-diff and code-diffs in the training corpus. 
Out of the code-diffs with the top \textit{k} scores, NNGen extracts commit messages based on the code-diff with the highest \bleu{} score.
Huang \etal{}~\cite{huang2020} proposed \textit{ChangeDoc}, a method for automatic commit message generation that utilizes existing commit messages in version control systems. 
ChangeDoc extracts similar messages by considering both semantic and syntactic information and uses this information to generate new commit messages.
Wang \etal{}~\cite{Wang2021} took a different path with their proposed solution called \textit{QAcom} where they automatically assess the quality of commit messages generated by various other approaches. This is achieved through retrieval techniques based on semantic relevance, which helps filter out poor-quality messages and retain high-quality ones.

Although promising, these approaches have limitations. They require a training dataset for reference, which can be limited in scope, leading to an out-of-vocabulary issue and potentially generating irrelevant or redundant messages. These limitations can result in poor quality and ineffective commit messages.

\noindent
\textbf{Hybrid Approaches:}
\textsc{ATOM}~\cite{liu2020}, incorporates an abstract syntax tree for representing code changes and integrates both retrieved and generated messages through hybrid ranking. Furthermore, Contextualized Code Representation Learning (CoreGen)~\cite{nie2021} leverages contextualized code representation learning strategies to improve the quality of commit messages. However, both approaches suffer from \textit{exposure bias}~\cite{ranzato2016} issues.
Wang et al. (2021b) proposed \textit{CoRec}, a hybrid model that combines generation-based and retrieval-based techniques. 
CoRec addresses the limitations of generation-based models, such as ignoring low-frequency words and exposure bias, by using an information retrieval module and decay sampling mechanism.
Shi \etal{}~\cite{shi2022} introduced \textsc{RACE}, a neural commit message generation model that incorporates retrieval-based techniques. \textsc{RACE} utilizes similar commits to guide the neural network model in generating informative and readable commit messages. 

The hybrid approaches discussed above have demonstrated impressive performance. However, they fail to account for the impact of a code change on the source code's control flow and data dependency and often neglect the context of the entire source code. 
Furthermore, their reliance on a retrieval module requires a training dataset during inference, which limits their scalability in real-world scenarios. This paper aims to provide a solution to address these identified limitations.

\section{Threats to Validity}
\noindent

\textbf{Construct validity} concerns with the degree to which our analyses measure
what we intend to analyze.
To ensure that our implementation works as intended,
we implemented automated tests and conducted manual code review for each module.
Prompt selection process for gpt-3.5-turbo and gpt-4-turbo models can be seen as a threat to validity
especially when the output of these models significantly depends on the provided prompts.
To mitigate this threat, we sought participation from multiple participants with prior software development to help us select the prompts. 
During the survey, we provided the participants with the results of all prompts, alongside the corresponding ground truth message for a particular code diff.
This helped to ensure that the prompts chosen were relevant, effective, and unbiased.

\textbf{Internal validity} is associated with the ability to draw conclusions from the performed experiments and results.
To minimize the associated threats with the implementation of compared techniques,
we attempted to use the replication package provided by the respective authors. 
In some cases, we did not find replication packages or were outdated.
In such cases, we contacted the authors of the paper for further clarification. 
Furthermore, we attempted to implement the approach ourselves if their replication package was not available or not reproducible.
Specifically, we implement CommitBert based on the information provided in the original publication.

\textbf{External validity} concerns with the ability to generalize the results.
Our study focuses on code changes to Java programming language.
However, \deltagraph{} and the rest of the implementation, by design,
works independently from the programming language.
To make our work reproducible, we make all the scripts, data, and developed tool available online~\cite{anonymous_github_2023, anonymous_2023_models}.

\section{Conclusion and Future Work}

In this work, we proposed a novel commit message generation approach \textsc{viz.} \comet{}.
The proposed approach leverages a new code change representation technique, 
\deltagraph{}, and a quality assurance module combined with a Transformer-based generator to achieve the goal of effective commit message generation.
Our results indicate that incorporating contextual information significantly improves the performance of encoder-decoder based models.
Based on that, we implemented our approach that considers contextual information.
We found that our approach achieves state-of-the-art performance compared to existing approaches.
Furthermore, we conducted a comparison between our approach and the gpt-3.5-turbo and gpt-4-turbo \api{} under zero-shot, one-shot, and multi-shot settings. The evaluation was performed against three prompts selected by participants in our survey.
Our approach demonstrated a significant performance advantage over gpt-3.5-turbo and gpt-4-turbo model.
In the future, we would like to explore how the quality assurance module can be used to address other subjective aspects of commit messages beyond those addressed in this work. 
Also, we aim to conduct an extensive study of multiple quality aspects that facilitate leveraging the purpose of commit messages and modify the training approach of the QA module to fit the findings accordingly. 
Furthermore, our focus will extend to enhancing the scalability of the trained model, enabling it to handle larger codebases automatically, and maintaining real-time representations, which will result in faster inference times.


\printbibliography

\if@endfloat\clearpage\processdelayedfloats\clearpage\fi






\end{document}